\documentclass[useAMS,usenatbib]{mn2e}
\usepackage{graphicx}
\usepackage{amssymb,amsmath}
\usepackage{times}

\title[V1200~Cen: a bright triple in the Hyades moving group]
{Orbital and physical parameters of eclipsing binaries from the 
All-Sky Automated Survey catalogue --- VII. V1200~Centauri: 
a bright triple in the Hyades moving group.\thanks{Based 
in part on observations collected through through CNTAC proposals 
CN-2012A-21, CN-2012N-36 and CN-2013A-93.}}
\author[J. Coronado et al.]{
J. Coronado$^1$\thanks{jccorona@astro.puc.cl}, 
K. G. He{\l}miniak$^{2,3,1}$, 
L. Vanzi$^{4,5}$, 
N. Espinoza$^1$, 
R. Brahm$^1$, 
A. Jord\'an$^{1,5,6}$, 
\newauthor
M. Catelan$^{1,5,6}$,
M. Ratajczak$^{3}$ 
and M. Konacki$^{3}$\\
$^1$Instituto de Astrof\'isica, Pontificia Universidad Cat\'olica de Chile, 
Av. Vicu\~na Mackenna 4860, Santiago, Chile\\
$^2$Subaru Telescope, National Astronomical Observatory of Japan, 650 North Aohoku Place, Hilo, HI 96720, USA\\
$^3$Nicolaus Copernicus Astronomical Center, Department of Astrophysics, ul. Rabia\'{n}ska 8, 87-100 Toru\'{n}, Poland\\
$^4$Department of Electrical Engineering, Pontificia Universidad Cat\'olica de Chile,
Av. Vicu\~na Mackenna 4860, Santiago, Chile\\
$^5$Centro de Astro-Ingenier\'ia, Pontificia Universidad Cat\'olica de Chile, Santiago, Chile\\
$^6$Millennium Institute of Astrophysics, Santiago, Chile}

\pagerange{\pageref{firstpage}--\pageref{lastpage}} \pubyear{2014}

\def\LaTeX{L\kern-.36em\raise.3ex\hbox{a}\kern-.15em
    T\kern-.1667em\lower.7ex\hbox{E}\kern-.125emX}
    
\begin{document}

\label{firstpage}

\maketitle

\begin{abstract}
We present the orbital and physical parameters of the detached 
eclipsing binary V1200~Centauri (ASAS~J135218-3837.3) 
from the analysis of spectroscopic 
observations and light curves from the \textit{All Sky Automated Survey}
(ASAS) and SuperWASP database. The radial velocities were computed 
from the high-resolution spectra obtained with the OUC 50-cm telescope 
and PUCHEROS spectrograph and with 1.2m Euler telescope and CORALIE 
spectrograph using the cross-correlation technique \textsc{todcor}. We found that 
the absolute parameters of the system are $M_1= 1.394\pm 0.030$ M$_\odot$, 
$M_2= 0.866\pm 0.015$ M$_\odot$, $R_1= 1.39\pm 0.15$ R$_\odot$, 
$R_2= 1.10\pm 0.25$ R$_\odot$. We investigated the evolutionary status and kinematics of 
the binary and our results indicate that V1200~Centauri is likely a member of the Hyades moving group, but the largely inflated secondary's radius may suggest that the system may be even younger, around 30 Myr. We also found that the eclipsing pair is orbited by another,
stellar-mass object on a 351-day orbit, which is unusually short for hierarchical triples. This makes V1200 Cen a potentially interesting target for testing the formation models of multiple stars.
 \end{abstract}


\section{Introduction}
Detached, double-lined spectroscopic binaries that are also eclipsing 
provide an accurate determination of stellar mass, radius and temperature 
for each of their individual components, and hence constitute a strong 
test of single star stellar evolution theory \citep{last}. Eclipsing 
binary (EB) stars are very important for the study of stellar astrophysics. 
Their particular geometrical layout, their dynamics and radiative physics 
enable a detailed and accurate modelling and analysis of the acquired data, 
and allow to measure many basic physical parameters of the components 
\citep{h2009}. Studies of binary stars by all the 
techniques available in modern astrophysics allow to measure a wide range of 
parameters for each of the component stars, with some of them determined with 
very high accuracy (e.g., uncertainties of less than 1\%).

This also refers to the pre-main-sequence (PMS) stars.
Through a complete analysis of spectroscopy and photometry of these systems 
orbital and physical parameters of the two stars can be accurately derived. 
However, until recently there were only seven known low-mass pre-main sequence EBs with 
$M$ $<$ 1.5 $M_\odot$: RXJ~0529.4+0041A \citep{cov2000,cov2004}, V1174~Ori 
\citep{stas2004}, 2MJ0535-05 \citep{stas2006,stas2007}, JW380 \citep{irwin}, 
Par~1802 \citep{cargi,stas2008}, ASAS~J0528+03 \citep{stempels}, and~MML53 
\citep{hebb}. 

Mutiple systems of two or more bodies seem to be the norm at all stellar evolutionary stages, according to observations, and it has been accepted for some time now that binarity and multiplicity is established as the principal channel of star formation \citep{Reip}. For this reason, knowledge of the formation of multiple stars becomes necessary to understand
star formation in general \citep{del2004}. Moreover, observations of young multiple systems allow to test evolutionary models in early stages of stellar evolution.


On the other hand, it is important to increase the sample of young stars (especially PMS) because main-sequence, 
solar-type stars are well described by stellar 
evolution models (observations agree well with theoretical isochrones), 
but recent measurements of the stellar properties of low-mass dwarfs and 
young PMS stars remain problematic for the existing models \citep{mo-ca2012}. 



In this paper we present the orbital and physical analysis of an eclipsing binary
V1200~Centauri (HD~120778, HIP~67712, ASAS~J135218-3837.3; $\alpha$~=~13:52:17.51, $\delta$ = -38:37:16.82).
It is a well detached system with a circular orbit, a short orbital period ($\sim$2.5 days),
and observed parameters consistent with the pre-main-sequence 
evolutionary stage. We also announce the discovery that it is a triple hierarchical 
system.

This system was first studied in 1954 in the Cape Photographic Catalog 1950 
\citep{jack}, but only the magnitudes and epoch. Then in 1984 the spectral type 
of F5V was defined \citep{houk}. The system is described as an Eclipsing Algol (EA) by 
\citet{otero} and has a $V$ magnitude of 8.551 \citep{and-fra}. 
As a bright star, it was a subject of spectroscopic studies
of the Geneva-Copenhagen survey \citep{Nordstr,holm2009}, where was treated 
as a single star, however due to large brightness ratio in the visual
we can consider some of their results reliable.
The parallax and proper motion are also known: $\pi$ = 8.43(94)~mas,
$\mu_\alpha$ = -72.49(87), $\mu_\delta$ = -44.20(78)~mas~yr$^{-1}$
\citep{vLe07}.

In Section 2 we describe the observations used to identify V1200~Cen as an 
eclipsing and spectroscopic binary with a circumbinary companion, 
in Section 3 we determine the orbital 
parameters of the system and physical properties of the component stars of 
and compare them with theoretical isochrones. 
Finally, in Section 4 we discuss the future observations needed 
to fully analyse the system.



\section{Observations}

\subsection{Echelle spectroscopy}
Observations with PUCHEROS instrument were carried out between 
January and May 2012 at the 50-cm telescope of the Observatory UC Santa 
Martina located near Santiago, Chile. The telescope is the European Southern 
Observatory (ESO) instrument, formerly located at La Silla. PUCHEROS 
is the Pontificia Universidad Cat\'olica High Echelle Resolution 
Optical Spectrograph developed at the Center of Astro-Engeneering of 
Pontificia Universidad Cat\'olica de Chile \citep{infante,vanzi}. 
The spectrograph is based on a classic echelle 
design, fed by the telescope through an optical fibre and it covers the 
visible range from 390 to 730 nm in one shot reaching a spectral resolution 
of 20,000. It is located in a stable position, in a room adjacent to the 
telescope. The science image was typically created from four separate 
observations, 15 minutes each, combined later into one frame. 
For the wavelength calibration we used exposures of ThAr lamps taken 
before and after the science sequence.
The data reduction of PUCHEROS spectra was performed with the task 
echelle in the IRAF package, following standard steps: CCD reduction,
spectrum extraction and wavelength calibration. 

The data from PUCHEROS were supplemented with spectra observed with the 
CORALIE spectrograph at the 1.2 m Euler telescope. CORALIE is a fibre-fed 
cross-dispersed echelle spectrograph, it covers the spectral range from 
380 to 690 nm reaching a spectral resolution of 50,000. Observations were 
made in a simultaneous wavelength calibration mode, where the light of the 
object is collected by one fibre, and of the ThAr lamp by the other. The 
spectra were taken between June 2012 and July 2013.
CORALIE data were reduced with a dedicated, Python-based pipeline 
\citep{jor14}.

\subsection{Photometry}
Photometric data were obtained from the All-Sky Automated Survey (ASAS) 
\citep{pojm,pacz} and from the SuperWASP transiting planet survey \citep{poll2006}. ASAS 
has produced an extensive catalogue of variable stars (ACVS) of the southern 
hemisphere\footnote{\texttt{http://www.astrouw.edu.pl/asas/?page=acvs}}. 
In this work, we use 495 data points from the 
third stage of the survey, obtained in the $V$ filter between 2000 and 
2009. SuperWASP is a wide-field photometric variability survey 
designed to detect transiting gas-giant planets around bright main 
sequence stars. The survey cameras observe bright stars (V$\sim$ 9-13) 
at high precision (1\%) using a broad $V+R$ band filter. 3234 
data points have been extracted from the SuperWASP public 
archive\footnote{\texttt{http://exoplanetarchive.ipac.caltech.edu/\\applications/ExoTables/search.html?dataset\\=superwasptimeseries}}.

\section{Analysis}

\subsection{Radial velocities and orbital solution}

\begin{table*}
\centering
\caption{Radial velocity (RV) measurements of V1200~Cen, together with the 
final measurement errors ($\sigma$) and residuals from the final three-body 
fit ($O-C$). All values in km~s$^{-1}$. In the last column, 5/P denotes 
OUC-50cm/PUCHEROS and E/C Euler 1.2m/CORALIE observations.}\label{tab_rv}
\begin{tabular}{lrrrrrrc}
\hline\hline
JD-2450000 & $v_1$ & $\sigma_1$ & $O-C_1$ &  $v_2$ & $\sigma_2$ & $O-C_2$ & Tel./Sp. \\
\hline
 5714.615861 &   45.958 & 0.646 & -0.340 & --- & --- & --- & 5/P \\
 5736.539995 &   64.395 & 0.494 &  0.222 & -127.519 & 2.640 & -0.757 & 5/P \\
 5737.639889 &  -67.029 & 0.523 &  0.394 &   88.377 & 4.198 & -1.718 & 5/P \\
 5750.604835 &  -62.826 & 2.446 &  1.482 & --- & --- & --- & 5/P \\
 5751.584224 &   74.651 & 0.536 & -0.788 & -126.370 & 4.324 &  3.856 & 5/P \\
 6066.642808 &   47.460 & 1.498 & -0.858 & --- & --- & --- & 5/P \\
 6066.665643 &   51.655 & 0.841 &  0.658 & --- & --- & --- & 5/P \\
 6078.565477 &  -36.335 & 2.129 &  1.846 & --- & --- & --- & 5/P \\
 6080.625298 &  -89.867 & 0.163 & -0.009 &  112.325 & 1.503 &  0.657 & E/C \\
 6081.564728 &   52.113 & 0.228 & -0.021 & -116.745 & 1.075 & -0.456 & E/C \\
 6179.474281 &  -26.024 & 0.167 &  0.035 &   80.336 & 0.885 & -1.105 & E/C \\
 6346.690592 &  -12.831 & 0.169 & -0.321 &   67.855 & 0.876 &  0.573 & E/C \\
 6348.857536 &  -55.020 & 0.165 &  0.255 &  136.192 & 1.064 &  1.044 & E/C \\
 6349.894755 &   94.865 & 0.194 &  0.023 & -107.687 & 1.017 & -0.499 & E/C \\
 6397.520928 &   38.353 & 0.112 & -0.034 &  -71.655 & 0.772 &  1.072 & E/C \\
 6398.517694 &  -77.575 & 0.116 &  0.017 &  112.000 & 0.951 & -0.331 & E/C \\
 6497.610599 &  -67.667 & 0.157 & -0.152 &  133.439 & 0.797 &  0.229 & E/C \\
 6498.610654 &   64.361 & 0.113 &  0.082 &  -78.099 & 0.942 &  0.453 & E/C \\
\hline
\end{tabular}
\end{table*}

\begin{figure}
\centering
\includegraphics[width=\columnwidth]{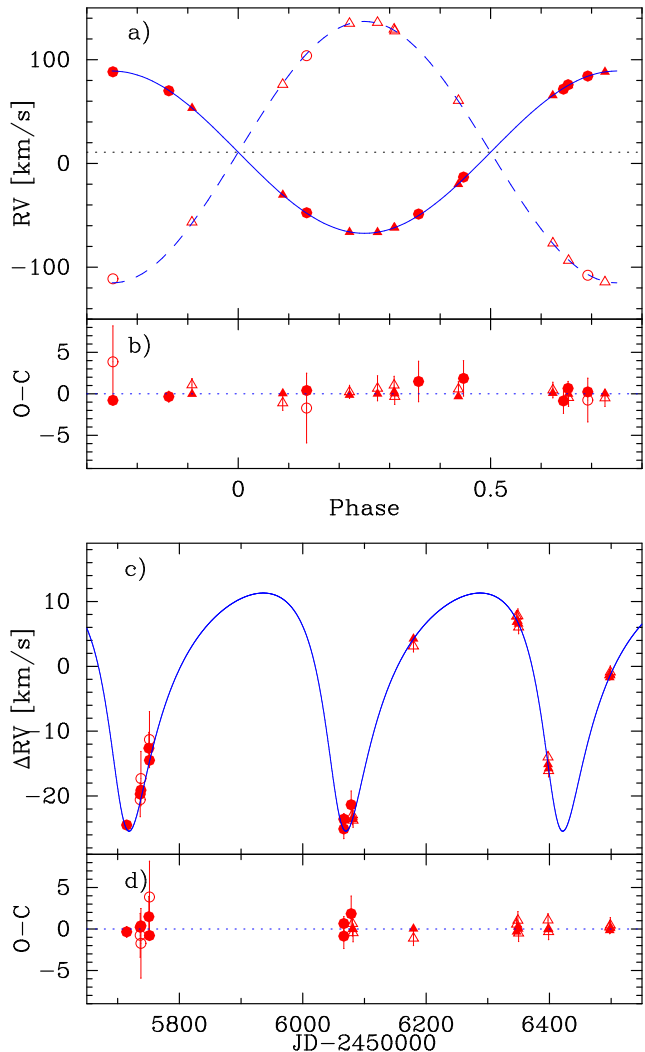}
\caption{Three-body orbital model of V1200~Cen based on the RV 
measurements from PUCHEROS (circles) and CORALIE (triangles). 
Filled symbols are for the primary component, and open ones for 
the secondary. Black dotted line on panel a) marks the systemic 
velocity $v_\gamma$. {\it a)} Keplerian orbit of the AB pair as 
a function of the orbital phase, with the perturbation removed; 
{\it b)} residuals of the full model as a function of phase; 
{\it c)} perturbation from the third body as a function of time, 
Keplerian orbit removed; {\it d)}~residuals of the full model 
as a function of time.}\label{fig_rv}
\end{figure}

Radial velocities (RVs) were measured with an implementation of the 
two-dimensional cross-correlation technique \textsc{todcor} \citep{zuc94} 
with synthetic spectra used as templates. The formal RV measurement 
errors were computed from the bootstrap analysis of \textsc{todcor} 
maps created by adding randomly selected single-order \textsc{todcor} 
maps. The peaks of the cross-correlation function (CCF) coming from 
both components were significantly different in height, due to the large 
brightness ratio of the two stars. The PUCHEROS spectra usually had 
lower signal-to-noise (S/N) ratio (5-30) and the CCF peak from the 
secondary was not always recognized. The S/N of the CORALIE data was 
normally higher (25-60) and the secondary's CCF peak 
was more prominent, still very low though.
With \textsc{todcor} we also tried to estimate the intensity 
ratio of the two components, but due to low S/N of the cool 
secondary the results were very uncertain and we find them unreliable.

The orbital solution was done simultaneously on all RV measurements 
with our simple procedure, which fits a double-Keplerian orbit using 
the Levenberg-Marquartd algorithm \citep[for a more detailed description, 
see][]{h2009}, and allows for Monte-Carlo and bootstrap analysis to 
obtain reliable estimations of uncertainties. We used the photometric 
data first to find the orbital period, and check for eventual 
eccentricity (see next Section). The eccentricity was found to be 
indistinguishable from zero so it was held fixed to 0.0 in the further steps. 
In the orbital analysis also the period and $T_0$ were held fixed. 
In the final orbital analysis we have also kept fixed the period and $T_0$. We the used values found by JKTEBOP (see Table 3), i.e:
\begin{equation}
Min_I = 2451883.8813(19) + E 2.4828752(22)
\end{equation}
The weight of the points was scaled according to the formal errors 
found by \textsc{todcor}. The procedure also allows for fitting for 
the differences of the zero points of different instruments, separately 
for each component. Those shifts were found to be of a similar value 
than their uncertainties, and different for each star. This can be 
explained by a mismatch between the spectrum and the template used, 
and large rotational velocities of the two stars. 

We started with a purely double-Keplerian solution 
with no perturbations, but the solutions we were getting were not 
satisfactory, with high reduced $\chi^2 \simeq 1600$, $rms$ of the 
fit about 10 km/s for both components, and residuals of both components 
correlated, i.e. differing from the model by a similar value. We then 
used a modified version of our procedure to look for a third, 
circumbinary body in the system. We fitted for the outer orbit's period 
$P_3$, amplitude of the inner pair velocity variations $K_{12}$, and 
the base epoch $T_3$, eccentricity $e_3$ and the argument of 
pericentre $\omega$. We ran the fitting procedure again on the whole data 
set, and found a satisfactory solution, characterised by a much lower 
$rms$ -- 0.98 and 1.50 km/s for the primary and secondary, respectively 
and much lower reduced $\chi^2 = 1.29$.  All of our spectra
have barycentric correction, we used IRAF's \textit{bcvcor} for the PUCHEROS data and in the case 
of CORALIE data the correction is implemented in the reduction pipeline. Therefore, the scale of the 
perturbation and the outer orbit's eccentricity 
can't be explained by the improper correction for the barycentre.
Hereafter, following the usual 
convention, we will refer to the eclipsing pair as AB, and to the third 
body as C. 

We present all the measurements, together with the errors and residuals, 
in Table \ref{tab_rv}. In Table \ref{tab_orbit} we present our 
results of the full orbital analysis. The parameters of the circumbinary 
orbit should be treated as preliminary, but the binary pair's orbital 
elements are well constrained. The RV measurement errors were initially 
scaled in such way that the final reduced $\chi^2$ was close to 1, and 
the fit itself was not affected (weights not changed). The errors were 
found by a bootstrap analysis (10000 iterations). In such a way we take care of the 
possible systematics and obtain reliable uncertainties of the final 
parameters. Our model, separated into the Keplerian and perturbation 
components, is presented in Figure \ref{fig_rv}. 

\begin{table}
\centering
\caption{Results of the RV analysis and the orbital parameters of V1200~Cen.}\label{tab_orbit}
\begin{tabular}{lcc}
\hline\hline
Parameter & Value & $\pm$\\
\hline
\multicolumn{3}{c}{\it AB eclipsing pair}\\
$K_1$ [km~s$^{-1}$] & 78.23 & 0.37 \\
$K_2$ [km~s$^{-1}$] & 126.0 & 1.1 \\
$v_\gamma$ [km~s$^{-1}$] & 10.92 & 0.94 \\
$a_{12} \sin{i}$ [R$_\odot$] & 10.026 & 0.058\\
$e_{12}$ & 0.0 & (fixed) \\
$q$ & 0.6208 &  0.0062\\
$M_1 \sin^3{i}$ [M$_\odot$] & 1.352 & 0.027 \\
$M_2 \sin^3{i}$ [M$_\odot$] & 0.839 & 0.012 \\
\multicolumn{3}{c}{\it Outer orbit}\\
$P_3$ [d] & 351.5 & 3.4 \\
$T_3$ [JD-2540000] & 5358.2 & 8.0 \\
$K_{12}$ [km~s$^{-1}$] & 18.37 & 0.11 \\
$e_3$ & 0.42 & 0.09 \\
$\omega_3$ & 156 & 7 \\
$a_3\sin{i_3}$ [AU] & 0.538 & 0.033\\
$f$ [M$_\odot$] & 0.099 & 0.025\\
$M_3 (i_3=90^\circ)$ [M$_\odot$] & 0.662 & 0.066 \\
\multicolumn{3}{c}{\it Other fit parameters}\\
E/C$-$5/P${_1}^a$ [km~s$^{-1}$] & 0.98 & 0.31 \\
E/C$-$5/P${_2}^a$ [km~s$^{-1}$] & 5.32 & 2.00 \\
DoF & \multicolumn{2}{c}{20} \\
$rms_1$ [km~s$^{-1}$] & \multicolumn{2}{c}{0.68}\\
$rms_1$ [km~s$^{-1}$] & \multicolumn{2}{c}{1.40}\\
\hline
\end{tabular}
\\$^a$ E/C$-$5/P is the difference in spectrograph zero 
\\points measured for each component separately. 
\end{table}

It is worth noting that the mass function $f(M_3)$ was calculated 
from the full formula:
\begin{equation}
f(M_3) = \frac{M_3^3 \sin^3{i_3}}{(M_1+M_2+M_3)^2}
\end{equation}
and is quite large. The minimum mass of the third body (for 
$i_3=90^\circ$) is a substantial fraction of the total mass of the 
eclipsing pair, so the popular approximation 
$(M_1+M_2+M_3)^2 \simeq (M_1+M_2)^2$ is not valid. We checked for 
tertiary eclipses in the residuals of the light curve models (see 
next Section) and we found no obvious evidence for them, which 
means that the outer orbit's inclination is different from 
$\sim90^\circ$. We have also failed to identify a tertiary peak in 
the CCF, which means that the tertiary component is significantly 
dimmer than the secondary. All this is consistent with for example 
a 0.7 M$_\odot$ star on a $70-75^\circ$ orbit. 
{It is also not excluded that the component C may itself be 
a binary composed of two similar, lower-mass stars, which together 
exceed the mass of B. This would be in agreement with \citet{tok14},
 who found no hierarchical triples with outer orbits shorter than 
$\sim$1000 days. The mechanism leading to a formation of such double-double 
system, with short inner orbital periods and outer period shorter 
than 1000$\sim$ d, assumes non-aligned outer and inner orbits 
\citep{whi01,tok14}. It is thus possible that in V1200~Cen the mutual 
inclination between the outer and inner orbit is large, making the 
Kozai mechanism possible (Kozai 1962)}.

\subsection{Light curve modelling}

\begin{figure}
  \centering  
  \includegraphics[width=\columnwidth]{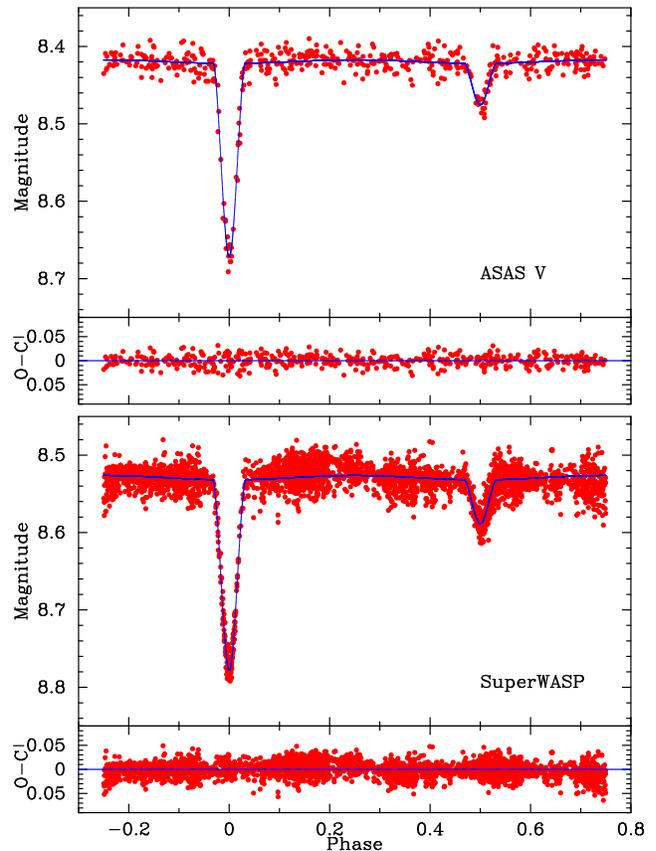}
  \caption{Observed ASAS (top) and SuperWASP (bottom) light curves 
with \textsc{jktebop} models over-plotted. Lower panels show the residuals of the fit.
}
  \label{fig_lc}
\end{figure}

\begin{table*}
\centering
\caption{Parameters obtained from the the ASAS and SuperWASP (SW) LC analysis.
The adopted values are weighted averages.}\label{tab_lc}
\begin{tabular}{lcccccc}
\hline\hline
Parameter & ASAS Value & $\pm$ &SW Value & $\pm$ & Adopted Value & $\pm$ \\
\hline 
    $P$ [d] & 2.4828778 & 0.0000043& 2.4828752 &  0.0000025  &2.4828752 & 0.0000022\\
    $T_{0}$ [JD] &1883.8789  &  0.0031 &1883.8827 &  0.0024 &1883.8813 & 0.0019 \\ 
    $i$ [$^\circ$]  & 81.9  & $^{+2.8}_{-1.3}$ &81.6 & $^{+1.6}_{-1.3}$ & 81.8 & $^{+1.4}_{-1.2}$\\
    $r_{1}$ & 0.137 & $^{+0.014}_{-0.015}$ & 0.138 & $^{+0.025}_{-0.034}$ & 0.137 & $^{+0.014}_{-0.015}$ \\
    $r_{2}$ & 0.107 & $^{+0.024}_{-0.039}$ & 0.110 & $^{+0.038}_{-0.026}$ & 0.109 & $^{+0.022}_{-0.025}$ \\
\hline
\end{tabular}
\end{table*}

   \begin{figure*}
   \centering
\includegraphics[width=0.8\textwidth]{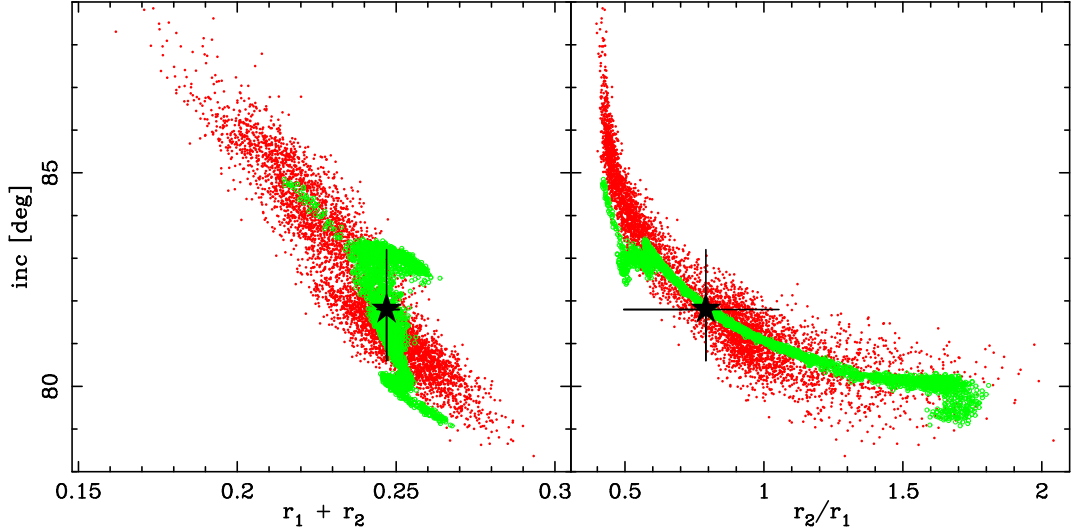}
   \caption{Results of the error analysis performed with \textsc{jktebop} with Monte-Carlo on the ASAS (red)
and residual shifts on the SuperWASP data (green). Plots present the distribution of consecutive solutions on the 
$r_1 + r_2$ vs. $i$ (left) and $k=r_2/r_1$ vs. $i$ (right) panels. Black stars with error bars correspond to the adopted values with their $1\sigma$ uncertainties.}
              \label{fig_inc_rr}
    \end{figure*}

Both ASAS and SuperWASP light curves were fitted using the \textsc{jktebop} 
code (Southworth et al. 2004a,b), which is based on the \textit{Eclipsing 
Binaries Orbit Program} \citep[\textsc{ebop};][]{pophet81}, and \textsc{phoebe} 
\citep{PZ2005} -- an implementation of the WD code \citep{wd71}.
\textsc{jktebop} determines the optimal model light curve that matches the 
observed photometry and reports the parameters obtained for the model. 
The parameters derived include the period $P$, time of minimum light $T_{0}$, 
surface brightness ratio, relative sum of the radii  $(R_{1} + R_{2})/a$, 
ratio of the radii $R_2/R1$, inclination $i$, eccentricity $e$,
and argument of periastron $\omega$. The routine also takes into account 
the effects of limb darkening, gravity darkening and reflection effects. 
We adopted the logarithmic limb darkening law from \cite{claret}. 
To obtain the limb darkening coefficients we considered the ASAS light 
curve to be in Johnson's $V$ passband. For the SuperWASP filter we 
interpolated the coefficients from \cite{poll2006}. {We also checked the ASAS data for O-C timing variations, and we did not find any significant ones. Thus we did not correct the light curves for the light time effect induced by the third body.}
We used \textsc{phoebe} to obtain the temperature of the secondary star, using the 
temperature of the primary star known previously from \citet{holm2009}. 
We did not use the latest Geneva-Copenhagen survey results,
as they rely on infrared data, which in the case of V1200~Cen can be
affected by the secondary and tertiary star.
The value of the mass ratio was found in the previous orbital analysis.
Figure \ref{fig_lc} shows the observed ASAS and SWASP light curves together 
with their models. 

Reliable uncertainties in case of ASAS data were calculated with the 
Monte-Carlo method (10000 runs) and with residual-shifts \citep{sou08} in case of 
SWASP data. Figure \ref{fig_inc_rr} shows the distribution of $r_1+r_2$ and $k$
as the function of the inclination angle. The usual correlation is clearly visible.
The results of the LC  analysis with \textsc{jktebop} are presented in Table \ref{tab_lc}.
We found that the ratios of radii ($k$) and effective temperatures 
are the most uncertain values, vastly contributing to errors of such physical parameters 
as absolute radii or luminosities. This is due to large scatter of the 
photometric data and low contribution of the star B (12\% in $V$). 
One way to constraint $k$ would be to use intensity ratios from \textsc{todcor},
but as we mentioned the S/N of the spectra was too low. Data of a much higher 
S/N are needed, optimally taken in IR (both photometry and spectroscopy), 
where the secondary's contribution is larger.


\subsection{Kinematics}
Using the known parallax and proper motion \citep{leeuwen} and our value 
of the systemic velocity $v_\gamma$, we have calculated the galactic
velocities: $U=-36.7\pm3.3$, $V=-21.8\pm3.6$ and $W=-1.8\pm0.6$~km~s$^{-1}$
(no correction for the solar movement has been done).
These values put V1200~Cen in the Hyades moving group \citep{sea07,zha09},
which suggests the age of $\sim$625~Myr. However, \citet{fam08} have shown
that about half of the stars that reside in the same area in the velocity
space as the Hyades group, actually does not belong to it. Nevertheless,
V1200~Cen seems to {be a young system belonging} to the thin galactic disk.

\subsection{Absolute parameters}

\begin{table}
\centering
\caption{Physical parameters of V1200~Cen obtained with \textsc{jktabsdim}
on the basis of spectroscopic values of $T_{\rm{eff,1}}$ and $[Fe/H]$.}\label{tab_phys}
\begin{tabular}{lcc}
\hline\hline
Parameter & Value & $\pm$ \\
\hline 
      $a$ [R$_\odot$] & 10.13 & $^{+0.07}_{-0.06}$\\
    $M_1$ [M$_\odot$] & 1.394 & 0.030 \\
    $M_2$ [M$_\odot$] & 0.866 & 0.015 \\
    $R_1$ [R$_\odot$] & 1.39  & $^{+0.14}_{-0.15}$\\
    $R_2$ [R$_\odot$] & 1.10  & $^{+0.22}_{-0.25}$\\
    log $g_{1}$ & 4.30 & $^{-0.09}_{+0.10}$\\
    log $g_{2}$ & 4.29 & $^{-0.18}_{+0.20}$\\
    $[Fe/H]$ & -0.18$^a$ &  \\
    $T_{\rm{eff,1}}$ [K] & 6266$^a$ &  94 \\ 
    $T_{\rm{eff,2}}$ [K] & 4650$^b$ & 900 \\
    $L_1$ [$\log($L/L$_\odot)$] &  0.42 & $^{+0.09}_{-0.10}$\\
    $L_2$ [$\log($L/L$_\odot)$] & -0.29 & $^{+0.38}_{-0.40}$\\
    $d$ [pc] & 98 & 11\\
\hline
\end{tabular}
\\$^a$ From \citet{holm2009}.
\\$^b$ From temperature ratio obtained with \textsc{phoebe}.
\end{table}

The absolute dimensions and distance were calculated using 
\textsc{jktabsdim} with the results obtained from the radial velocity 
and light curve analysis. The parameters with their respective 
uncertainties used for the input were the velocity semi amplitudes 
(km~s$^{-1}$), period (days), orbital inclination (degrees), 
fractional stellar radii (i.e. in units of the orbital major semi-axis), 
effective temperatures, and apparent magnitudes in the $B$ and $V$ filters. 
We did not use the available $JHK$ photometry as it could have been 
affected by the third star. For the interstellar reddening we considered E(B-V)=0, 
and we observed no major differences for the distances obtained in the two filters, 
both in agreement with the value of the parallax 8.43$\pm$0.94~mas
\citep[119$\pm$13~pc;][]{leeuwen}. We present the the physical
parameters of V1200~Cen in Table \ref{tab_phys}.

\subsection{Evolutionary status}

\begin{figure*}
 \begin{picture}(500,600)
  \put(0,0){\includegraphics[width=0.45\textwidth]{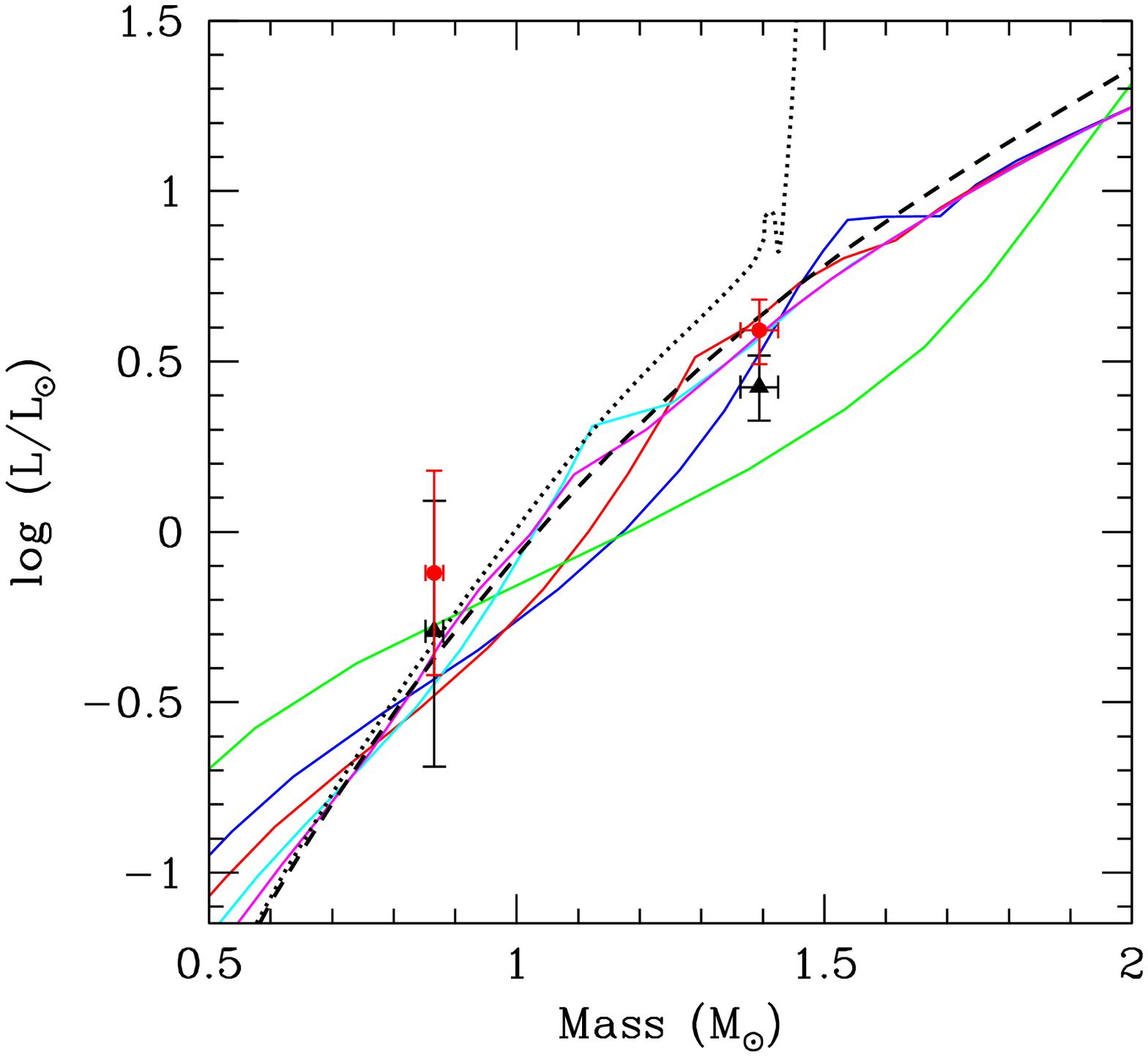}}
  \put(250,0){\includegraphics[width=0.45\textwidth]{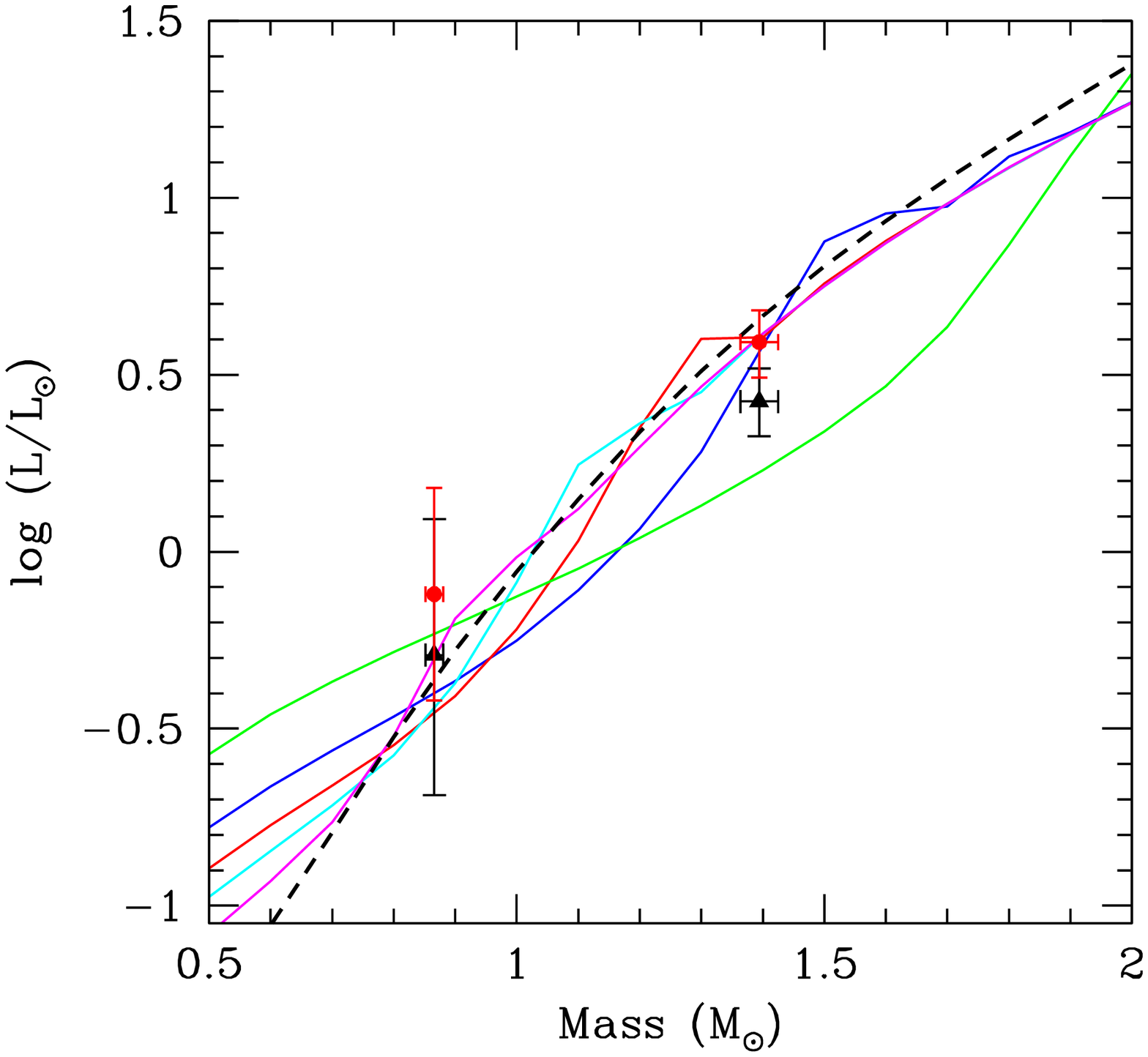}}
  \put(0,200){\includegraphics[width=0.45\textwidth]{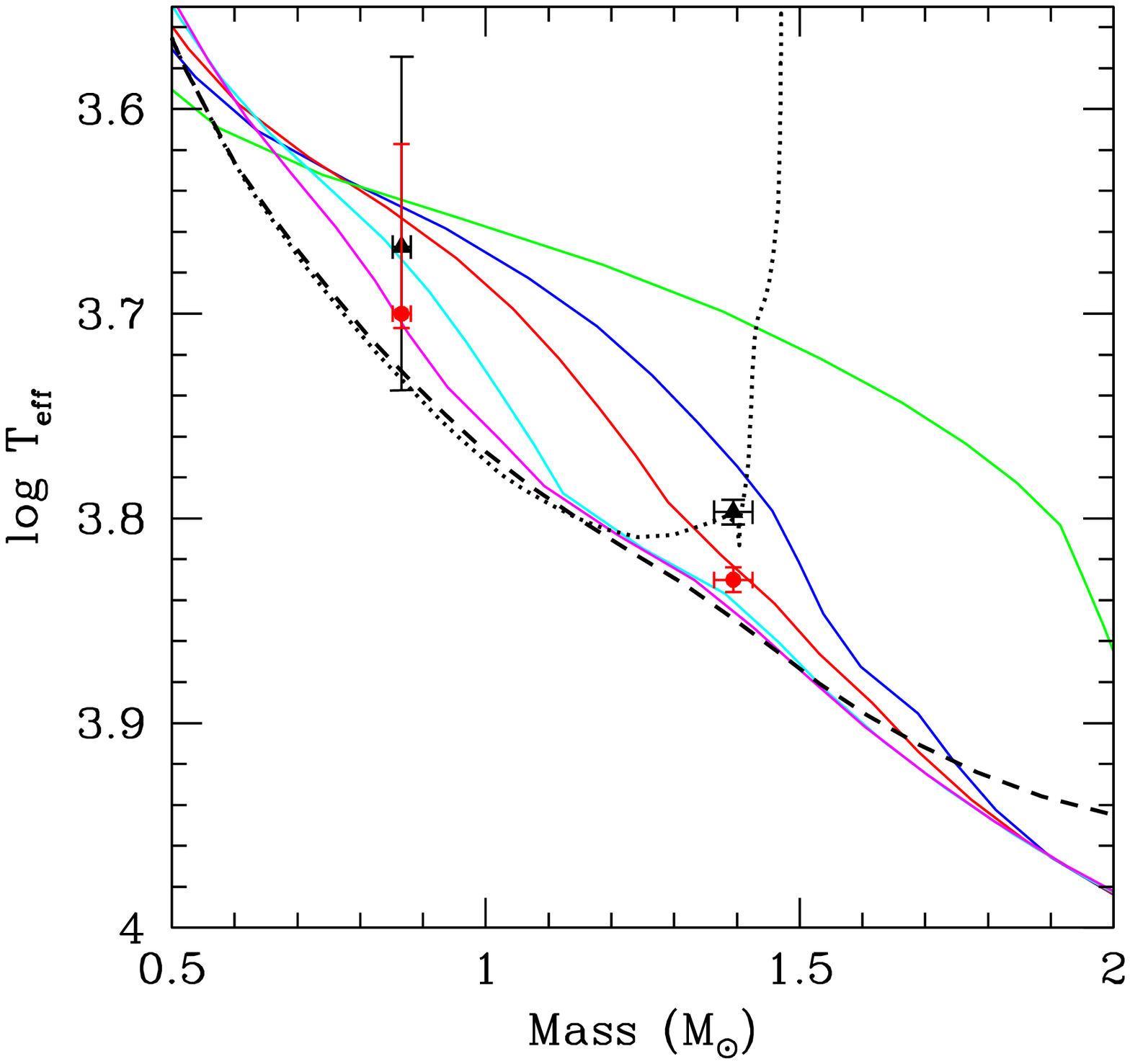}}
  \put(250,200){\includegraphics[width=0.45\textwidth]{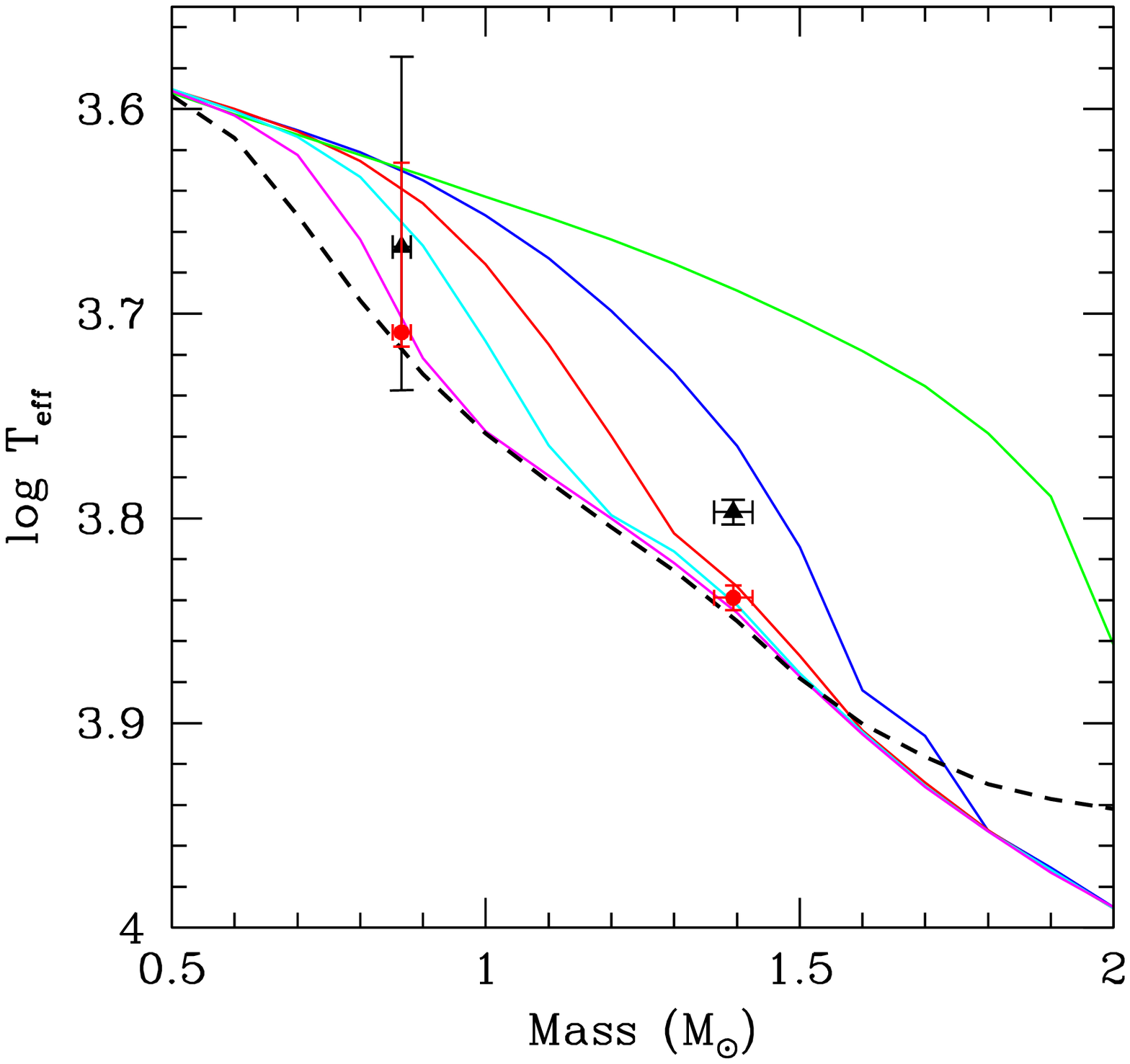}}
  \put(0,400){\includegraphics[width=0.45\textwidth]{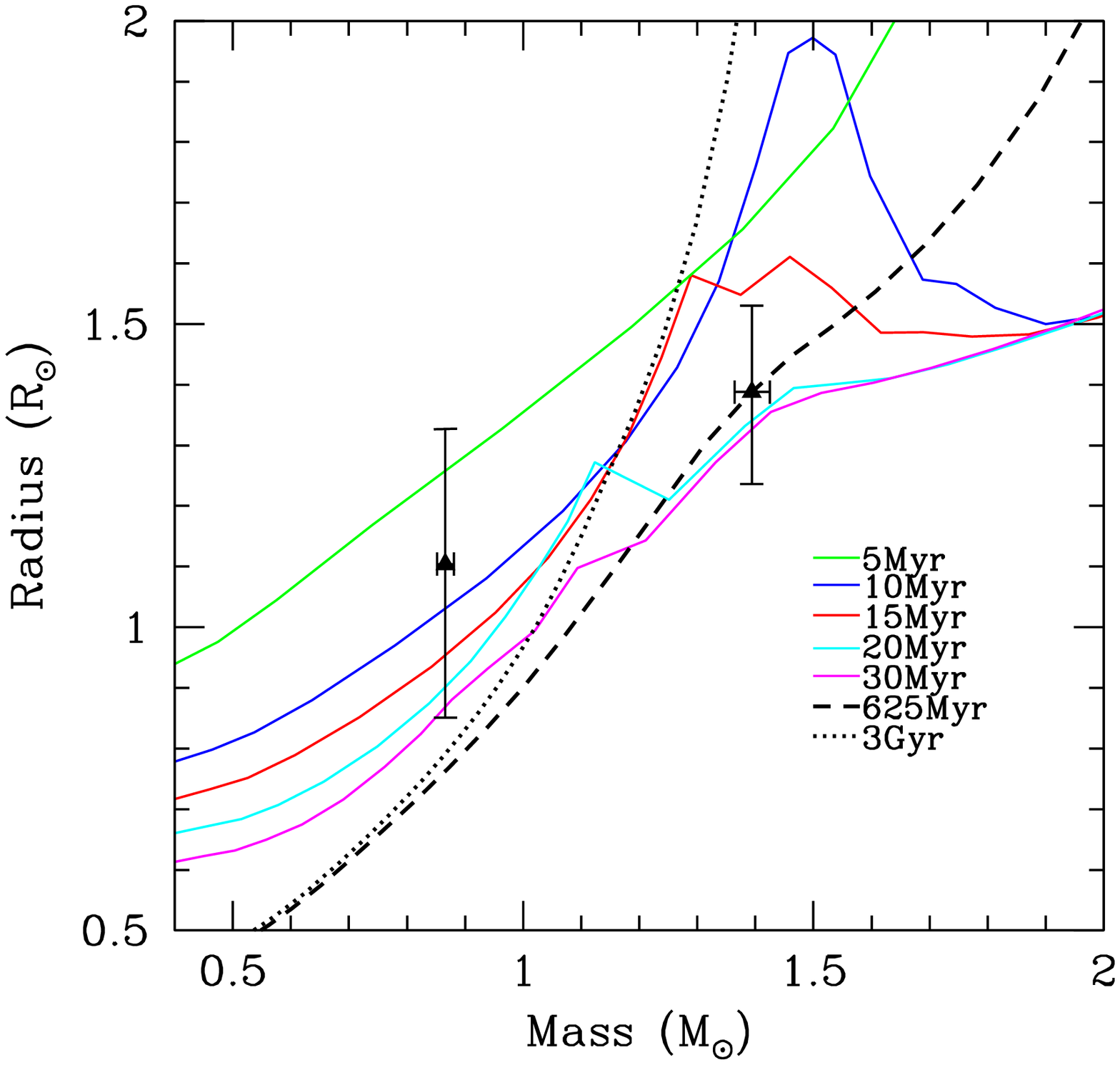}}
  \put(250,400){\includegraphics[width=0.45\textwidth]{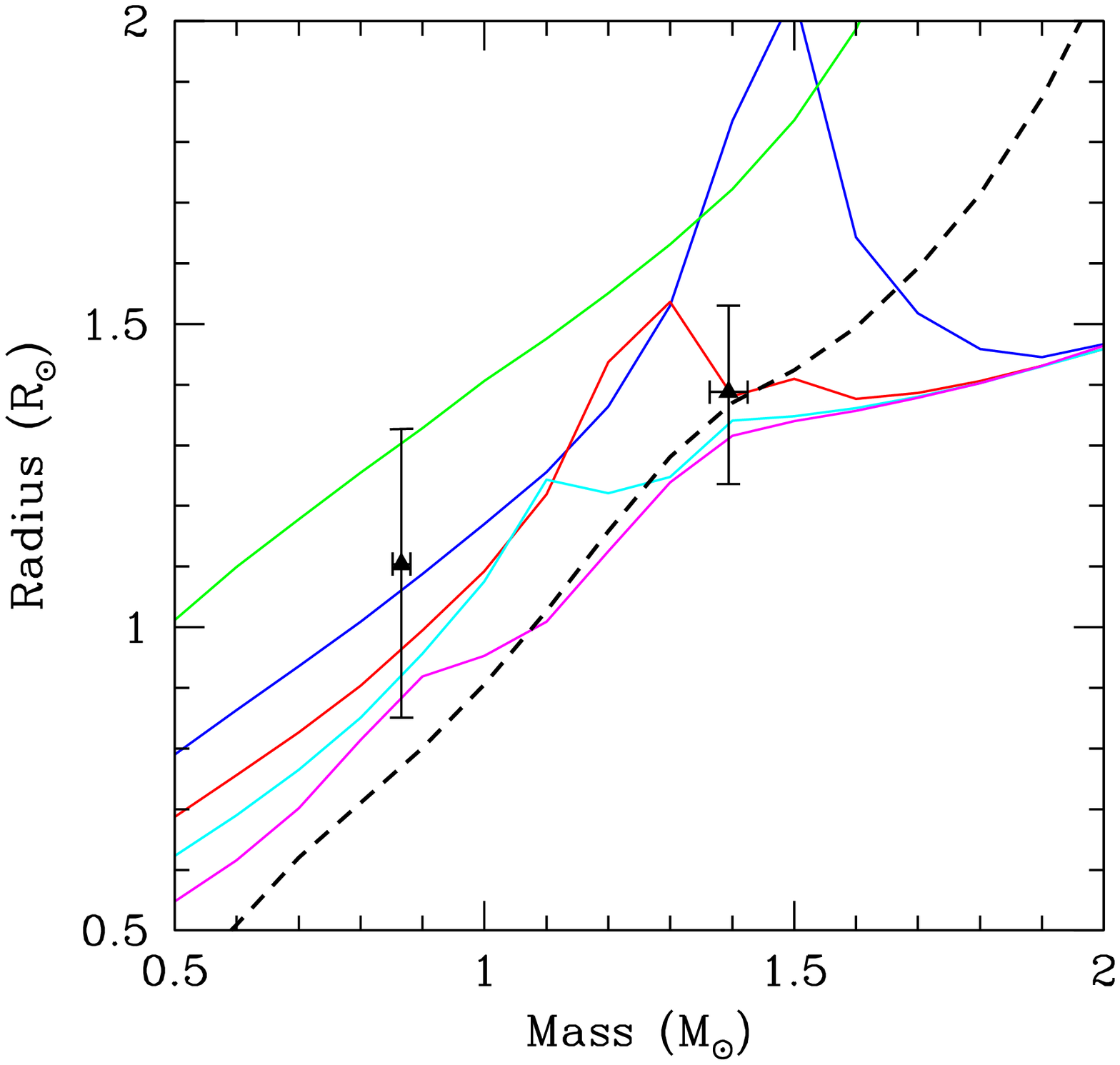}}
 \end{picture}
  \caption{Mass vs. radius, $\log(L)$ and log($T_{\mbox{eff}}$). The left 
column shows our results and the isochrones from 
Y$^{2}$ for ages of 6, 10, 15, 20, 625 Myr and 3 Gyr. The right column shows
the same but using the \citet{sies} isochrones (without 3 Gyr).
{ Black} symbols are for our results from Table \ref{tab_phys} and { red}
symbols for the results from Table \ref{tab_phys_hot}.}\label{fig_iso}
\end{figure*}

In Figure \ref{fig_iso} we present our results from Table \ref{tab_phys}
plotted over theoretical stellar evolution models of \citet{sies} and from Yonsei-Yale 
\citep[Y$^{2}$;][]{yi2001} set for ages of 6, 10, 15, 20 and 625 Myr.
For the Y$^{2}$ we also include the 3 Gyr isochrone.

The values from Table \ref{tab_phys} are marked as black points.
The two most precise values we have are the mass and temperature of the 
primary. One can see that it is too cool for a main sequence object, so it is 
either evolved, as the 3~Gyr isochrone suggests, or at its pre-main-sequence 
(PMS) stage. However, both Y$^2$ and Siess' models predict a drastic change in temperature 
and radius between 10 and 20 Myr. The primary's radius is much too small for an 
evolved star and agrees with the late PMS, meanwhile the temperature clearly suggests
younger age, inconsistent with the radius. We also note that secondary's parameters
are much better reproduced by PMS models, especially its radius, which is much 
larger than main-sequence objects of the same mass, as expected for 
stars that are still evolving onto the main sequence.

The comparison of our results with the model predictions indicates that V1200~Cen
is a pre-main-sequence system, but no age is fully consistent with the data,
and the resulting distance is only in a fair agreement with results from the 
$Hipparcos$. Unfortunately, the uncertainties are very large, especially for the 
secondary component, so our conclusions cannot be treated as final. The strongest 
constrain we have comes from the primary's temperature, which was derived 
spectroscopically by \citet{holm2009}. { We find almost exactly the same 
temperature -- 6263~K -- using the observed $B-V$ colour (0.475~mag) and 
calibrations by \citet{sek2000}. One has to remember that the Holmberg's 
temperature was derived under the assumption that the star is 
single.} Thus, it is { still possible} that the true $T_{eff}$ is higher. 
We run a series of tests to find the temperatures that give the best agreement with 
the $Hipparcos$ parallax, assuming their ratio to be the same as found by us with 
\textsc{phoebe}. { We found the best match to the observed distance for much higher 
temperatures of 6900 and 5120~K for the primary and secondary respectively.}
The resulting effective temperatures and related luminosities are summarised in 
Table \ref{tab_phys_hot}. { We plot them on the Figure \ref{fig_iso} with red symbols.}

\begin{table}
\centering
\caption{Radiative parameters of V1200~Cen obtained with \textsc{jktabsdim}
by fitting the temperature scale to match the $Hipparcos$ distance.}\label{tab_phys_hot}
\begin{tabular}{lcc}
\hline\hline
Parameter & Value & $\pm$ \\
\hline 
    $T_{\rm{eff,1}}$ [K] & 6900 & 100$^a$ \\ 
    $T_{\rm{eff,2}}$ [K] & 5120$^b$ & 900$^a$ \\
    $L_1$ [$\log($L/L$_\odot)$] &  0.59 & $^{+0.09}_{-0.10}$\\
    $L_2$ [$\log($L/L$_\odot)$] & -0.13 & $^{+0.36}_{-0.37}$\\
\hline
\end{tabular}
\\$^a$ Uncertainty assumed.
\\$^b$ From temperature ratio obtained with \textsc{phoebe}.

\end{table}

One can see that the new { higher} values of radiative parameters fit the 
main sequence or even 20~Myr models, and both radii still agree with the 30~Myr 
isochrone. The age of 30~Myr also happens to be the time scale of circularisation of 
the system's orbit. All in all, we get a self-consistent model of a 
{ 30-625~Myr old, young multiple. As an attempt to distinguish between high and
low temperature scale we also run a simplified spectral analysis with the 
Spectroscopy Made Easy package \citep[SME;][]{val96}. We took the CORALIE spectra,
shifted by the measured velocity of the primary and stacked them together. We run the
SME on a portion of spectrum spanning 6190-6260\AA, and keeping the [$Fe/H$], $\log(g)$ 
and $v_{rot}$ on values from or expected from Tab. \ref{tab_phys}. The secondary's
contribution was treated as a contribution to the continuum, constant across 
the wavelength range. We obtained $T_{eff,1}\sim6000$~K, favouring the lower 
temperature scale, and younger ages (Fig. \ref{fig_iso}). We are aware of the 
fact that such analysis is affected by the secondary, and we do not treat it 
as a proof for the correctness of Holmberg's temperature, but we find it unlikely 
to be off by almost 1000~K, between spectral types F0 and F8.}
 

\section{Conclusions}
We report the discovery that the eclipsing binary V1200~Centauri 
is a triple, {likely a member of the Hyades moving group, but the largely inflated secondary's radius 
may suggest that the system may be PMS around 30 Myr}. There are few 
such objects known so far, however they are very important for 
calibrating stellar evolution models at young ages, where stars 
are changing rapidly as they evolve onto the main sequence. 
Analysis of ASAS and SuperWASP light curves combined with radial 
velocity measurements allowed to obtain absolute parameters of 
the system. Through our analysis of radial velocities and orbital solution we determined 
the presence of a third companion, in a wide orbit. We reached a good 
precision in mass determination (2.2 and 1.7\%), but other parameters 
(radii, temperatures, luminosities) are not that well established. 
Further analysis of the system allowed us to 
compare our results with stellar evolution models, obtaining an approximate 
age of 30 Myr. Despite its brightness, data of higher S/N are required to
better constrain physical parameters of the system, especially temperatures and
ratio of the radii. Possible solution would be to obtain photometry and 
spectroscopy in the IR, where components B and C contribute more than in 
visual. The secondary eclipse would be deeper, and the influence of the 
third light should be detectable. Also, if RVs of the star C were measured, 
a full dynamical solution of the system could be obtained, including mass 
of the third star and the inclination of its orbit. This would also
allow for comparing the isochrones with three stars, thus constrain
the age and evolutionary status even better.
With its probable age, V1200~Cen is an important object to study 
the tidal and third-body interactions young binaries.

\newpage
\section*{Acknowledgements}
PUCHEROS was funded by CONICYT through project FONDECYT No. 1095187. 
J.C acknowledges J.M. Fern\'andez for his assistance during all of the observations with PUCHEROS. K.G.H. acknowledges support provided by the by the National 
Astronomical Observatory of Japan as Subaru Astronomical Research 
Fellow and the Proyecto FONDECYT Postdoctoral No. 3120153. L.V acknowledges support by Fondecyt 1095187, 1130849 and FONDEF CA13I10203.
N.E and R.B are supported by CONICYT-PCHA/Doctorado Nacional.
We also acknowledge the support provided by the Polish National Science 
Center through grants 2011/03/N/ST9/01819, 2011/01/N/ST9/02209 and 5813/B/H03/2011/40. Support for A.J. and M.C. 
is provided by the Ministry for the Economy, Development, and Tourism's Programa Iniciativa Cient\'{i}fica Milenio 
through grant IC\,120009, awarded to the Millennium Institute of Astrophysics (MAS), and by Proyecto Basal PFB-06/2007. M.C. acknowledges additional support by FONDECYT grant \#1141141.


\label{lastpage}

\end{document}